\documentclass[conference, 10pt]{IEEEtran}

\usepackage{amsmath,amsfonts}
\usepackage{algorithmic}
\usepackage{algorithm}
\usepackage{graphicx}
\usepackage{textcomp}
\usepackage{threeparttable}
\usepackage{xcolor}
\usepackage{multirow}
\usepackage{pifont}

\makeatother

\newcommand{\revision}[1]{\textcolor{black}{#1}}
\newcommand{\new}[1]{\textcolor{black}{#1}}

\newcommand{\iccad}[1]{\textcolor{black}{#1}}

\begin{document}

	\bstctlcite{IEEEexample:BSTcontrol}
	
	\title{Optimizing Malware Detection in IoT Networks: Leveraging Resource-Aware Distributed Computing for Enhanced Security }

	\author{\normalsize
		\begin{tabular}[t]{ccc}
			\large Sreenitha Kasarapu & \large Sanket Shukla & \large Sai Manoj Pudukotai Dinakarrao \\
			Dept. of ECE  & Dept. of ECE & Dept. of ECE\\
			George Mason University & George Mason University & George Mason University\\
			Virginia, USA & Virginia, USA & Virginia, USA\\
			e-mail: skasarap@gmu.edu & e-mail: sshukla4@gmu.edu & e-mail: spudukot@gmu.edu\\
			\vspace{-10pt}
		\end{tabular} \vspace{-2em}}
	
	\maketitle

	\begin{abstract}

		In recent years, networked IoT systems have revolutionized connectivity, portability, and functionality, offering a myriad of advantages. However, these systems are increasingly targeted by adversaries due to inherent security vulnerabilities and limited computational and storage resources. Malicious applications, commonly known as malware, pose a significant threat to IoT devices and networks. While numerous malware detection techniques have been proposed, existing approaches often overlook the resource constraints inherent in IoT environments, assuming abundant resources for detection tasks. This oversight is compounded by ongoing workloads such as sensing and on-device computations, further diminishing available resources for malware detection. To address these challenges, we present a novel resource- and workload-aware malware detection framework integrated with distributed computing for IoT networks. Our approach begins by analyzing available resources for malware detection using a lightweight regression model. Depending on resource availability, ongoing workload executions, and communication costs, the malware detection task is dynamically allocated either on-device or offloaded to neighboring IoT nodes with sufficient resources. To safeguard data integrity and user privacy, rather than transferring the entire malware detection task, the classifier is partitioned and distributed across multiple nodes, and subsequently integrated at the parent node for comprehensive malware detection. Experimental analysis demonstrates the efficacy of our proposed technique, achieving a remarkable speed-up of 9.8$\times$ compared to on-device inference, while maintaining a high malware detection accuracy of 96.7\%.
	\end{abstract}

	
	


	
	\vspace{-0.5em}
	\section{Introduction}
	\vspace{-0.5em}
	

	The substantial increase of research in the field of Internet of Things(IoT), has lead to massive developments. New innovations are making it easy to connect day-to-day use gadgets to internet. Innovations such as smart homes, smart watches, smart cars, etc., have become an essential part of human life.
	
	With their huge data handlings, IoT networks become a desirable target for cyber attackers who prey on user information.The prominent cyber-attacks are due to malware. Malware is a malicious software developed to infect a network to explore and steal information such as passwords, financial data, and manipulating the stored data without the user's consent. Each year IoT networks suffer many security threats, only in the year of 2020, there were more that 5 billion recorded malware attacks. The technical advancements, lead adversaries to create new complex malware, it is recorded that 8 million and more malware threats are detected every day. 
	
	The predominant increase in malware attacks and security threats, fuel up the growing concerns of IoT device's safety. Techniques which can detect malware in IoT devices and stop user data exploitation, are needed.
	Despite effectiveness, the existing works suffer from three primary challenges: 
	\textbf{(1) Real-Time Malware Detection:} 
	Malware detection techniques can be categorized into two categories: 
	i) static analysis, and ii) dynamic analysis. 
	Unlike static analysis \cite{static_limits}, which analyses the internal structures of malware binaries, dynamic analysis \cite{dynamic} is a functionality test that makes it effective in identifying the presence of malware in an application.
	Recent works on malware detection (both static and dynamic analysis)  techniques utilize a variety of Machine Learning (ML) techniques to enhance the performance \cite{nataraj}. 
	Among the ML-based malware detection techniques, the CNN-based image classification technique \cite{cnn_detect} is observed to be efficient due to its prime ability to learn image features.
	However, it is important to detect malware during runtime 
	with minimal latency, as malware can propagate from one device to another before it gets detected. 
	The existing static analysis \cite{sta_dy} and dynamic analysis \cite{dynamic} techniques can detect the 
	malware; the real-time detection cannot be guaranteed due to the involved computations. 

	\revision{\textbf{(2) Reliable Feature Extraction:} 
		In addition, the reliability of malware detection depends on the features that 
		are captured and analyzed. Recent works have shown that the captured features may not be effective or reliable due to the issues in the sensing mechanisms. These reliability issues can range from simple over-counting to misrepresenting system-level features (or traits) as application-level features \cite{Das'19}. 
		\revision{For instance, hardware performance counters (HPCs) information is widely used in dynamic analysis for malware detection \cite{HPC_FCN}. 
			However, 
			utilizing HPCs' information for security has raised some concerns in recent years \cite{Das'19}, especially in terms of the reliability of the obtained HPC information. 
			For instance, the `Instruction count' in the Intel Pentium 4 is often over-counted \cite{Das'19}. Further, the coexistence of applications influences the HPC values and trends, 
			leading to non-determinism and unreliability. 
			Advanced malware crafting techniques such as code-obfuscation, metamorphism, and polymorphism \cite{morphism}  further increase the challenges on the reliability of observed and analyzed features.}  
	}
	
	
	
	Another pivotal challenge with malware detection on IoT devices is 
	\textbf{(3) Limited Resources in IoT Devices:} IoT devices are designed with limited resources for portability and performing a limited set of operations \cite{IOT_ps}. 
	A majority of the resources in such IoT devices are employed for executing user applications and are often resource-constrained and contentious. 
	However, the existing works assume that all the resources are available to perform security analysis, i.e., malware 
	detection. 
	Thus, there is a dire need to perform malware detection co-located with the existing 
	applications and meet the timing and reliability requirements. 


	This work addresses the aforementioned limitations by a novel resource-aware and workload-aware model parallelism-inspired malware detection technique for mid-scale IoT devices. 
	The proposed approach employs HPC-based on-device malware detection, which is further enhanced with model parallelism approach 
	to enable efficient malware detection even under stringent resource constraints. Furthermore, 
	the process of HPC collection is refined for robustness and efficiency. 
	The proposed approach performs efficient malware classification without demanding excessive on-device resources. 
	The novel contributions of this work can be outlined as follows: 
	
	\begin{itemize}
		\item We introduce a robust HPC collection mechanism for effective input in security applications. 
		\item To enable resource awareness, a lightweight automatic resource assessment is incubated on IoT devices. 
		Based on the available resources, offloading the malware detection task is determined. For this purpose, 
		a lightweight regressor is developed that utilizes the workload of the IoT devices and ML model parameters as variables.
		\item A resource-aware model parallelism-based ML model distribution to neighboring devices 
		for malware detection is designed. HPC traces of a given application are collected during runtime and are validated through a traditional ML classifier for malware detection.
	\end{itemize}
	
		
		

	
	The IoT devices focused on in this work are mid-scale with basic data processing ability, such as Jetson Nano boards.
	The experimental results prove that the proposed resource-aware model parallelism technique can detect complex malware in IoT networks with an accuracy $>$90\%. 
	Experimental analysis shows that the proposed technique can achieve a speed-up of 9.8$\times$ compared to on-device inference while maintaining a malware detection accuracy of 96.7\%.

	The rest of the paper is organized as follows: Section \ref{related} describes the related work and its shortcomings and comparison with the proposed model. Section \ref{problem} describes the problem for malware detection in IoT devices. Section \ref{proposed_sol} describes the proposed architecture resource-aware model parallelism, which assists with efficient malware detection in IoT devices, using a distributed runtime model training methodology. The experimental evaluation of the proposed model and comparison with various ML architectures is illustrated in Section \ref{results} and followed by the conclusions drawn from the paper are furnished in Section \ref{conclusion}.

	\vspace{-0.5em}
	\section{State-of-the-Art}
	\label{related}
	\vspace{-0.5em}
	
	\begin{figure*}[htbp]
		\centering
		\includegraphics[width=7in]{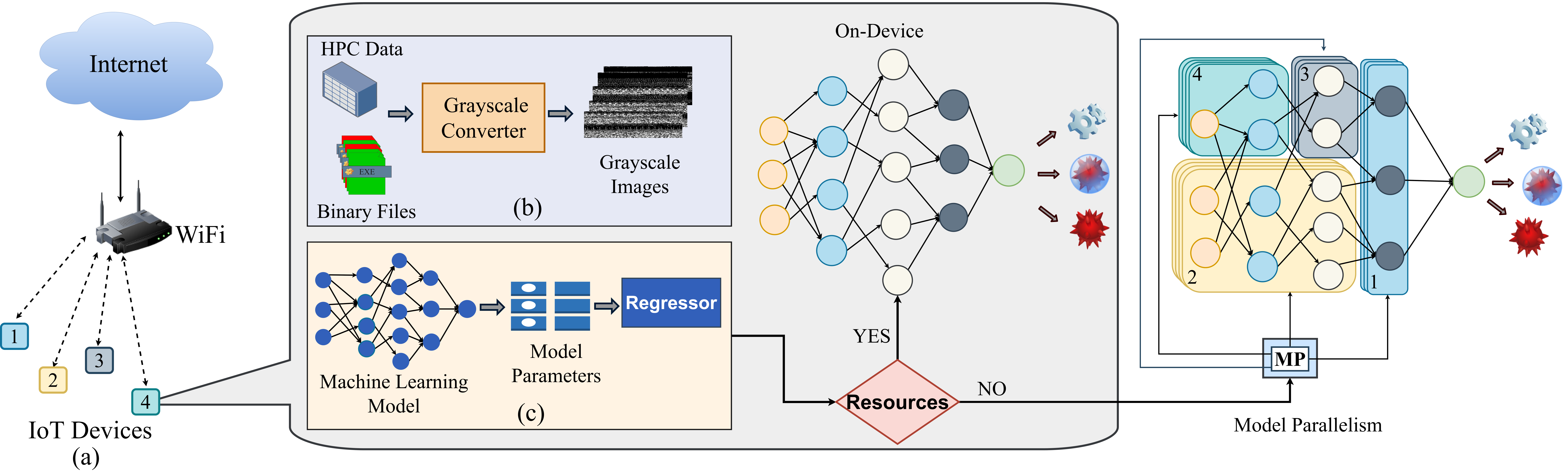}
		\vspace{-1.0em}
		\caption{(a) Distributed IoT Devices Framework, (b) HPC and Binary Data Pre-processing to Extract Input Image Dataset, (c) Framework to Identify the Resources in the Model} 
	
\vspace{-1.5em}
\label{fig:overview}
\end{figure*}
Resource-constrained IoT networks naturally become target to adversaries looking to attack on vulnerable systems. Due to their memory limitations they cant enforce strong defence against security threats. To address the security needs, techniques which can detect malware must be provided to these edge devices. Traditionally, static and dynamic analysis of malware detection are employed. Static analysis detects the malware by analyzing the internal structure of the binary executable files. This technique is performed in a non-runtime environment, as it doesn't need any executions. Is it a basic malware detection technique and often doesn't be reliable. It is only used for its simple and quick solution.

Dynamic analysis is a malware detection technique, performed in a secured runtime environment, like Sandbox. It is a functionality test and the binary files are executed to detect malware functionalities in them. Dynamic analysis is much more promising than static analysis in  malware detection. But it is not efficient in detecting complex malware families and hidden malware code blocks which do not execute in sandbox. 

With the increase in Machine Learning popularity, techniques were introduced for malware detection \cite{sanket_abhijitt_date2021, sanket_cases_2019, sanket_dac_2021, sanket_date_2023, sanket_dhavlle2021novel, sanket_glsvlsi_2022, sanket_iccd_2022, sanket_icmla_2019, sanket_ictai_2019, sanket_isqed_2024, sanket_rram_glsvlsi_21, sreenitha_kasarapu2021demography, sreenitha_mdpi, sreenitha_sanket_ubol, sreenitha_sathwika_vlsid, raghul_iscas}. These techniques convert binary application files and convert them into grayscale images. The grayscale images are used as input to train ML algorithms. In works used ML algorithms such as Support Vector Machines (SVM) and  K-Nearest Neighbour for image classification. These techniques serve as a stepping stone but are unable to the level of efficiency needed. These ML algorithms lack the ability to capture detail malware code patterns in grayscale images. Artificial Neural networks were further employed to solve this problem, neurons present in ANNs work good with capturing image patterns. But they have fully connected layers in network architecture, these layers consume extensive resources. Later, Convolutional Neural Networks(CNNs) were employed for this task \cite{sreenitha_sanket_aspdac, sreenitha_sanket_glsvlsi, sreenitha_sanket_tcad}, they can efficiently work with complex image data using feature extraction of Convolutional 2D layer. The input parameters can then be down-sampled using Maxpooling 2D layer. Thus, serving as an efficient image classification algorithm with lesser resource consumption. Still, IoT devices cant accommodate the memory requirements of the CNNs.

In addition to malware detection, we present some works on model parallelism. 
Model parallelism \cite{Verbraeken'20} is a technique where each node has the same data, but the ML model is divided. 
Model parallelism is suitable for training a massive ML when there are limited resources \cite{raghul_RS2020, raghul_SaravananICDSMLA'19, raghul_SaravananICICNIS'21, raghul_trng2020, Raghul2019}.    
In \cite{algebraic}, authors propose linear-algebraic-based model parallelism for deep learning networks. This framework allows the parallel distribution of any tensor in the DNN. Model parallelism is also mainly used in natural language processing. In \cite{megatron}, authors train a multi-billion parameter-based transformer language model. With the aid of multiple GPU nodes and pipeline structures, they could train such a gigantic model. 
In \cite{3d_transformers}, authors build a 3-dimensional distributed model to accelerate the training in the language model. They use a 3D model to complement matrix multiplication and vector operations in the transformer models. The use of model parallelism in security applications is not much explored.
To the best of our knowledge, this is the first work that employs model parallelism for the purpose of malware detection.


\section{Problem Formulation}
\label{problem}
\vspace{-0.25em}

With technology advancements, attackers are introducing complex hidden malware, by sneaking them into general applications used by IoT devices. 
One can define the problem of reliable malware detection in resource-constrained IoT devices with minimum processing capabilities as follows:

\begin{equation}
\begin{aligned}
	& & \mathbb{C}(D^n): {X} \rightarrow {Y} \hskip 0.75em \\
	& \text{s.t.} 
	& \mathfrak{Res}[\mathbb{C}] < \mathfrak{Res}[node]
	\label{eq2}
\end{aligned}
\end{equation}

Where $\mathbb{C}$ is the classifier model trained with dataset $D^n$ to perform malware detection. The dataset $D^n$ contains, a combination of malware and benign  samples. After training, the classifier $\mathbb{C}$ will be able to classify any sample $X$ and map it to either malware class ${M}$ or benign class $B$. The output class is represented as $Y$. 
\iccad{But, the resources required to perform inference, represented as $\mathfrak{Res}[\mathbb{C}]$ should be less than the available resources in an IoT node, represented 
as $\mathfrak{Res}[node]$.}
\revision{If the constraint in equation \eqref{eq2} is not met, then the inference task cannot be carried out by the device. Our proposed technique solves this by introducing a novel resource-aware malware detection model by offloading the workload inference to neighboring nodes.}

\section{Proposed Resource- and Workload-aware Malware Detection}
\label{proposed_sol}

\subsection{Overview of the Proposed Technique}
\vspace{-0.25em}
\new{The overview of the proposed technique is shown in Figure \ref{fig:overview}. 
Figure \ref{fig:overview}(a) represents the IoT devices present in a network. 
Figure \ref{fig:overview}(b) describes the data collection process. 
HPC traces are considered as one of the inputs for the proposed technique to improve the reliability of malware detection. Along with the HPC data,
the binary files (malware and benign) are considered as inputs for malware detection. 
The HPC data and binary files are integrated and converted to grayscale images. These image samples are trained (offline) using machine learning 
algorithms 
(CNN) for effective malware detection.} As shown in Figure \ref{fig:overview}(c), an automatic resource estimation is performed using a lightweight regression model to analyze the resources required for malware detection. Depending on the available resources, executing workloads, and communication overheads, malware detection (inference task) is either performed on-device or off-loaded to neighboring nodes with sufficient resources, as shown in Figure \ref{fig:overview}. 
We provide the details of the proposed technique below. 
\begin{table*}[htb!]
\centering
\caption{Parameter Estimations per Each Layer in a CNN Algorithm }
\vspace{-0.7em}
\label{tab1}
\scalebox{0.75}
{
\begin{tabular}{|c|c|c|}
	\hline
	
	\textbf{Layers} & \textbf{Description} & \textbf{Parameters} \\
	\hline
	Input  & No learnable parameters  & 0  \\ 
	\hline
	CONV  & (width of filter * height of filter * No. of filters in previous layer+1) * No. of filters in current layer & $f_{conv} = (w*h*p) + 1)*c$   \\ 
	\hline
	POOL & No learnable parameters & 0  \\ 
	\hline
	FC & (current layer neurons * previous layer neurons)+1 * current layer neurons & $f_{FC} = (n_c * n_p) + 1*n_c$ \\ 
	\hline
	Softmax & (current layer neurons * previous layer neurons) + 1 * current layer neurons & $f_{S} = (n_c * n_p) + 1*n_c$ \\ 
	
	\hline
\end{tabular}
}
\vspace{-1.0em}
\end{table*}

\subsection{Pre-processing and Data
Collection}
\vspace{-0.25em}





The microarchitectural event traces captured through HPCs are necessary for malware detection.
To address the reliability concerns that exist with the state-of-the-art HPC collection techniques, 
we propose fine-tuning state-of-the-art model-specific registers (MSRs) available in the modern computing system architectures, which are the source of the HPC information. 
Secondly, to solve the non-determinism challenge in HPCs, we redesign HPC capturing protocols with 
proper context switching and handling performance monitoring interrupt (PMI) units in the system while collecting HPCs. To obtain the HPCs solely for a given application, context switching needs to be accommodated, thereby eliminating the contamination of the obtained HPCs. From our preliminary analysis, the overhead (in terms of latency) to perform context switching for MiBench applications is around 3\% of an average application runtime which is 
affordable for enhanced security. Furthermore, PMI ensures proper context switching and reading of HPCs. It has been seen that configuring PMI per process often leads to better capturing of the 
HPCs \cite{Das'19}. Through this two-pronged utilization of context-switching + PMI, we collect reliable HPCs. To address the challenges such as overcounting \cite{Das'19}, 
we perform calibration through testing.


In addition to the reliability of the HPC data, the limited number of available HPC registers compared to the microarchitectural events 
pose a significant challenge \cite{}. 
To perform real-time malware detection, one needs to determine the non-trivial microarchitectural events that could be captured through the limited number of HPCs and yield high detection performance. 
To achieve this, we use principal component analysis (PCA) for feature/event reduction on all the microarchitectural event traces
captured offline by iteratively executing the application. 
Based on the PCA, we determine the most prominent events and monitor them during runtime. The ranking of the events is determined as follows: 
\vspace{-0.5em}
\begin{equation}
\rho_i = \frac{cov(App_i,Z_i)}{\sqrt{var(App_i) \times var(Z_i)}}
\label{pca_eq}
\vspace{-0.5em}
\end{equation}
where $\rho_i$ is the Pearson correlation coefficient of any $i^{th}$ application. $App_i$ is any $i^{th}$ incoming application. $Z_i$ is an
output data containing different classes, backdoor, rootkit, Trojan, virus, and worm in our case. $cov(App_i, Z_i)$ measures covariance between
input and output. $var(App_i)$ and $var(Z_i)$ measure variance of both input and output data respectively. 
We select the most prominent HPCs based on the ranking and monitor them during runtime for efficient malware detection. 
These reduced features collected at runtime are provided as input to ML classifiers which determine the malware class label (${\widehat{Y}}$ $\Rightarrow$ Backdoor, Rootkit, Trojan, Virus and Worm) with higher confidence. 

\iccad{In this work, we capture the functionality of dynamic HPC attributes (values) and convert them as grayscale images. 
Capturing the range of HPC values for a particular executable (benign or malware) illustrates the trend in variation of the HPC values for benign and malware samples. Hence, we have unique patterns in grayscale images for each executable file. However, it should be noted that the grayscale images of the same class of malware tend to show similar texture in some portions of the grayscale image.
With the emerging complex malware families and adversaries employing techniques such as morphism, it is becoming hard to classify malware using traditional malware detection techniques such as MLP and DNN. As MLPs are not deep enough to identify complex patterns and DNNs containing multiple dense layers are bulky to employ. Convolutional Neural Networks (CNNs) \cite{cnn_detect} have been proven to work much better in finding patterns, even if spatially separated, such as polymorphed malware. Thus, the input HPC and binary images are fed to CNN for improved malware detection capability.} 

\subsection{Automatic Resource Estimation}

With the huge amount of generated data, a CNN model is trained (offline), which can detect malware. The inference task (using the pre-trained CNN model) should be performed on IoT devices to detect any malware in them. And often, the inference task can be huge to fit in a single IoT node due to the parameters involved. The number of resources available in each node changes based on its workload. Instead of manually calculating the parameters of CNN and estimating whether available resources on a node will be sufficient each time, a regression model is constructed. The binary regression model is trained using data such as CNN's parameters, memory requirements of these parameters, and available memory at each node. As output, the binary regression model gives an estimation of whether the CNN model can be trained on a single node or must be distributed onto multiple nodes.

\begin{algorithm}
\caption{Lightweight Linear Regression Algorithm}
\label{algo2}
\begin{algorithmic} 

\STATE  \hskip -1em {\textbf{Require}}: $B_{exe}$ (Benign application files), $M_{exe}$ (Malware application files)
\STATE  \hskip -1em {\textbf{Input}}: $\mathfrak{Res}[node], \mathbb{C}$


\STATE \textbf{define} $Regressor(\mathbb{C})$:

\STATE \hskip 1em \textbf{for} {$layer \leftarrow \mathbb{C}$}: \textbf{do}
\STATE \hskip 2em $ var \leftarrow f(W, B, A)$

\STATE \hskip 2em \textbf{if} ${layer \rightarrow CONV}$
\STATE \hskip 3em  $ par = f_{conv}(var)$
\STATE \hskip 2em \textbf{elif} ${layer \rightarrow  (FC \vee Softmax)}$ 
\STATE \hskip 3em $par = f_{FC}(var)$
\STATE \hskip 2em \textbf{else}
\STATE \hskip 3em $par = 0$
\STATE \hskip 2em {\textbf{end if}}

\STATE \hskip 2em $ \Bar(P).append(par)$ 

\STATE \hskip 1em \textbf{end for}

\STATE \hskip 2em $ \mathfrak{Res}[model] \leftarrow N*batch\_size*\Bar(P)*1KB$ 
\STATE \hskip 2em $X_R.features \leftarrow \{ W, A, B, \Bar(P), \mathfrak{Res}[model],$
\STATE \hskip 9em $\mathfrak{Res}[node] \}$
\STATE \hskip 2em $ Res \leftarrow \mathbb{R}:(X_R, \beta)$
\STATE \hskip 2em \textbf{return} $Res$

\end{algorithmic}

\end{algorithm}

As shown in algorithm \ref{algo2}, the binary regression algorithm is constructed (offline). The training features of the regressor are parameters of the CNN. So the parameters of each layer are calculated. For each layer of CNN $\mathbb{C}$, the variables weight matrix $W$, bias $B$, activation $A$ are collected and stored in the variable $var$. These variables contribute to parameter calculations of different layers in CNN. As shown in Table \ref{tab1}, input layer and pooling layer represented as $POOL$ of the Convolutional Neural Networks does not have any learnable parameters. So, parameters $par$ are zero for these two layers. For convolutional layer $CONV$, fully connected layer $FC$ and softmax layer  $Softmax$, the parameters are calculated using the equations shown in Table \ref{tab1}.

\iccad{If there are multiple CNN layers, the parameters are calculated multiple times with different activation functions $A$. At last, the estimated parameters of each layer are appended to give $\bar(P)$. Then, the resources required for the model $\mathfrak{Res}[model]$ are calculated. The resources needed are a function of parameters $\bar(P)$ for each batch of the N number of batches. For example, it can be assumed that each parameter needs one Kilo Byte (1KB) of resources for inference, based on which the final resource required is estimated to be in MBs ($\sim 5MB$). $X_R.features$ represents the features to be given as input to the regressor $\mathbb{R}$, which predicts the resource estimations $Res$. The features in $X_R.features$ include weight matrix $W$, bias $B$, activation $A$, parameters of CNN at each layer $\Bar(P)$, resource estimation of model $\mathfrak{Res}[model]$ and resource available at each node $\mathfrak{Res}[node]$.}

\vspace{-0.5em}
\subsection{Workload- and Resource-Aware Malware Detection}
\vspace{-0.25em}

\new{
Depending on the outcome of the workload estimator, malware detection (inference) is performed either on-device or distributed across the nodes. 
This addresses two main challenges: a) efficient malware detection depending on the available on-device resources; and b) partitioning reduces the 
inference time \cite{Verbraeken'20}. Even in the case of on-device detection, as the resources available are sufficient, timing constraints (latency) can be met to 
achieve real-time malware detection. In the case of offloaded malware detection, as the child nodes do not have complete information, 
the security and privacy of the information (task) are ensured. 
The partitioning is performed based on the independency of the nodes of the ML classifier, 
represented as a graph, and the workload that could be accommodated on the parent and child devices \cite{Geng'19}. We provide an upper bound on the number of devices to which the task can be distributed. } 


Techniques such as DistBelief \cite{Jeffrey'12} are highly dependent on the partitioning of the model. Thus, they can lead to varied performances in our case and hence not adaptable. 
We adapt AllReduce \cite{Pitch'09} paradigm in this project, where the parent node accumulates the gradients from the children nodes.
Partitioning the model across a cluster is transparent and requires no structural modifications. 
To update the gradients and perform other computations including inference, 
a synchronous Allreduce approach is utilized for better scalability \cite{Pitch'09}. 
However, a direct adaptation of such technique makes it vulnerable to faults such as the unavailability of data or garbage data from one device 
can stagnate or contaminate the whole process. 
In addition, the malicious behavior of the child node can lead to 
misclassification, which has high possibility in our application. 

To address such concerns, 
we deploy Downpour stochastic gradient descent (SGD) \cite{Duchi'11}. Downpour SGD is resilient to machine failures and data manipulations, 
as it allows the training and inferring to continue even if some model replicas are offline. 
\iccad{Each child node is made available with required gradients ready to be transferred. The parent node is given access to all the gradients to ensure uninterrupted parameter transfer between nodes and there will be no stalling because of node unavailability.
In this manner, we address the unavailability or reliability concerns of other devices in an IoT network for malware detection.}
It needs to be noted that the training happens offline, and inference is performed in real-time. 
To minimize the communication overheads, we let the parent device choose the child devices within a threshold radius $R$ for which the communication costs are lower and ensure the devices have a smaller workload. As frequent communication between parent and child nodes lead to large overheads,  
we let the system communicate whenever a device's output is required as input for another device. 

In contrast to the existing works on malware detection, this project devises a mechanism to reliably obtain different node-level features, and perform a including microarchitectural event patterns through HPCs, connectivity, and physical features. Further, a novel resource- and workload-aware distributed malware detection with model parallelism.






\begin{algorithm}
\caption{Pseudo-Code for Distributed Runtime Modelling of Malware Detection}
\label{algo3}
\begin{algorithmic} 
\STATE  \hskip -1em {\textbf{Require}}: $M$ (Malware grayscale images), $B$ (Generated Benign grayscale images)
\STATE  \hskip -1em {\textbf{Input}}: $D^n = \{B + M\}, \mathfrak{Mem}[model] $


\STATE \textbf{define} $Distribute\_CNN\_model()$:
\STATE \hskip 1em \textbf{for} {$ n \leftarrow range(0,x)$}: \textbf{do} 
\STATE \hskip 2em \textbf{if} $\mathfrak{Mem}[model] \leq \mathfrak{Mem}[node]$
\STATE \hskip 3em $ node.append(n)$
\STATE \hskip 3em $ \mathfrak{Mem}[node] \leftarrow \mathfrak{Mem}[0] + \mathfrak{Mem}[1]+...+\mathfrak{Mem}[n]$

\STATE \hskip 2em {\textbf{end if}}

\STATE \hskip 1em\textbf{end for}

\STATE \hskip 1em $l_n = nn.layern.cuda(n) $
\STATE \hskip 1em $ model = nn.Sequential(l_1, l_2,...,l_n)$
\STATE \hskip 1em $input = D^n.cuda(0) $
\STATE \hskip 1em $ output = model(input) $

\STATE \textbf{return} $O_m$
\end{algorithmic}
\end{algorithm}

Algorithm \ref{algo3}, represents the Pseudo-code for proposed distributed runtime-based modeling of malware detection. The function to distribute CNN, represented as $Distribute\_CNN\_model()$, is called based on the output of the regressor (if resources are not available). It takes the memory estimation $\mathfrak{Mem}[model]$ of the CNN model for malware detection as input. It compares the model memory $\mathfrak{Mem}[model]$ and available memory at each node $\mathfrak{Mem}[node]$. It appends multiple node memory elements to find the number of nodes, required to distribute the model. The number of nodes $n$ should have a combined memory more than or equal to the model memory $\mathfrak{Mem}[model]$. If this condition is met, the CNN is distributed on $n$ nodes. The different layers of the malware detection model $l_1, l_2,....., l_n$ are divided on $n$ and trained. The input data is made available to the input layers, by passing them to the $node0$. Communication between the nodes is made possible due to the interdependent variables and back pass algorithm by the function $ model = nn.Sequential(l_1, l_2,...,l_n)$. 
\new{This resource estimation provides a prediction of whether the inference can be performed on a single node or if it needs to be done using parallelism.} 

\vspace{-0.5em}
\section{Results}
\label{results}
\vspace{-0.5em}
\subsection{Experimental Setup}
\vspace{-0.25em}

For the IoT network setup, we deployed 20 IoT nodes encompassing Broadcom BCM2711, Jeston Nanos, and quad-core Cortex-A72 (ARM v8) 64-bit boards. These nodes are connected through a wireless interface (WiFi). The deployed
Jetson Nanos contains a 128-core NVIDIA Maxwell architecture-based GPU and Quad-core ARM® A57 CPU. Four JetsonNano boards
are employed for model parallelism, and they provide access to 
multiple CPU and GPU nodes. We have obtained malware and benign applications from VirusTotal \cite{rvirus} with 12,500 benign samples and 70,000 malware samples encompassing five malware classes: backdoor, rootkit, trojan, virus, and worm.

Using the Quick HPC tool, we execute these malware and benign applications in a sandbox environment and capture a range of HPC values (e.g., for 20 ns, 40 ns). These HPC attributes of benign and malware samples are converted to grayscale images of size 256 x 256. In this image dataset, 70\% of the data is divided into the training set and 30\% as the test set. A CNN is built on the training set (offline), and the inference is performed on Jetson Nano boards (online). 
\iccad{The CNN is built with four convolutional layers, each containing a (3 $\times$ 3) filter for feature extraction. The input images are reshaped into (32 $\times$ 32), as the input images are grayscale and have one channel; the input size for these convolutional layers is (32, 32, 1). The convolutional layers are followed by pooling layers. The data is then flattened to pass it to the dense layers. All the layers of the CNN architecture are built using a \textit{`relu'} activation function, except the dense output layer, which is equipped with a \textit{`softmax'} activation to produce probabilities for each class. The class with the highest probability is classified to be the output class. The performance of the CNN in distributed setting using model parallelism is presented in Table \ref{tab3}.}

\vspace{-0.5em}
\subsection{Impact on Latency with Proposed Technique}
\begin{figure}
\vspace{-6.0em}
\includegraphics[width=0.55\textwidth, height= 9cm]{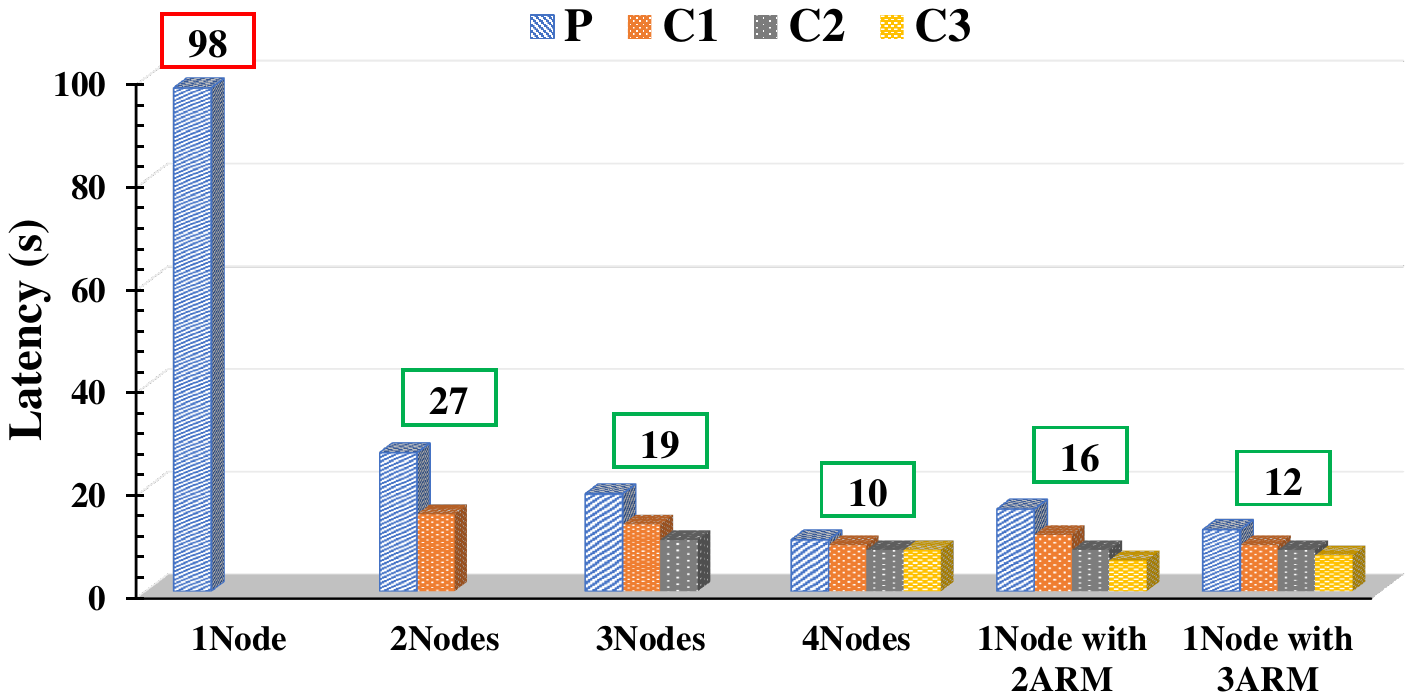}
\vspace{-7.5em}
\caption{Latency of Distributed learning for Malware Detection} 
\label{fig:latency}
\vspace{-1.5em}
\end{figure}

\vspace{-0.25em}
The inference task is done using the malware detection model, a convolution neural network algorithm trained on malware and benign samples. In a node, if the resources needed to perform inference are not enough, the malware detection task is off-loaded to multiple nodes. Normalized inference execution time is analyzed for cases a) the parent node has sufficient resources; b) the parent node does not have enough resources and outsources to multiple nodes. Figure \ref{fig:latency}, represent the latency of these cases. In Figure \ref{fig:latency}, Node represents the GPU on 
the Jetson Nano boards, and the ARM represents the Quad-core ARM® A57 CPU present in Jetson boards. Also, P represents the parent node, C1 represents the first child node, C2 represents the second child node, and C3 represents the third child node.
We observe that the inference time decreases with an increase in the number of nodes. As the parameters are divided over various nodes, the execution time needed for inference decreases. 
As the executions run in parallel, the total latency to perform inference in model parallelism is the latency of the node which takes the highest time to execute (usually, the model parallelism latency is the latency of parent node P). In Figure \ref{fig:latency}, for the case of sufficient resources, it takes 98 seconds to perform the inference task. 
For the case of model parallelism, we can observe a speedup of 4 $\times$ when the inference task is parallelized between two nodes. If we further off-load the inference task to three nodes, an additional speedup of about 1.5$\times$ is observed. 
The ARM boards used as child nodes in model parallelism also produced notable speed-ups. We observed an overall speedup of 9.8$\times$ while using four Jetson Nano boards. 

%



\vspace{-0.75em}
\subsection{Comparison with State-of-the-Art Works}
Table \ref{tab3} presents the comparison of the proposed technique with the existing HPC-based malware detection techniques. We compare the performance of the proposed technique in terms of accuracy, F1 score, and recall. All the models in table \ref{tab3} focus on malware detection based on HPC runtime features. Compared to the existing techniques, the proposed resource-aware CNN-based distributed training on HPC-based image data achieves the highest accuracy. \iccad{It maintains an average accuracy of 96.7\%. The average F1-score and recall show that the proposed technique achieves state-of-the-art HPC-based malware detection accuracy.}

\begin{table}[htb!]
\centering
\caption{Comparison with existing HPC-based detection techniques}
\vspace{-1em}
\label{tab3}
\scalebox{0.82}{
\begin{tabular}{|c|c|c|c|c|c|}
\hline

\textbf{Model} &  \textbf{Accuracy} & \textbf{F1-score} & \textbf{Recall} \\
& (\%) & & \\
\hline 
OneR \cite{HPC_r1} & 0.81  & 0.81 & 0.82 \\
\hline
JRIP \cite{HPC_r1} & 0.83 & 0.83 & 0.84  \\
\hline
PART \cite{HPC_r1}& 0.81 & 0.815 & 0.831 \\
\hline 
J48 \cite{HPC_r1} & 0.82  & 0.82 & 0.82  \\
\hline
Adaptive-HMD \cite{adaptive_hmd} & 0.853 & 0.853 & 0.858 \\
\hline
SVM \cite{results_nn}& 0.739  & 0.736 & 0.772  \\
\hline
RF \cite{results_nn} & 0.835 & 0.834 & 0.822 \\
\hline
NN \cite{results_nn} & 0.811 & 0.811 & 0.816 \\
\hline
SMO \cite{results_smo} & 0.932 & 0.933 & 0.931 \\
\hline
\textbf{Proposed (MP)} & \textbf{0.967} & \textbf{0.967} & \textbf{0.972} \\
\hline

\end{tabular}}
\end{table}


\begin{figure}
\includegraphics[width=\linewidth]{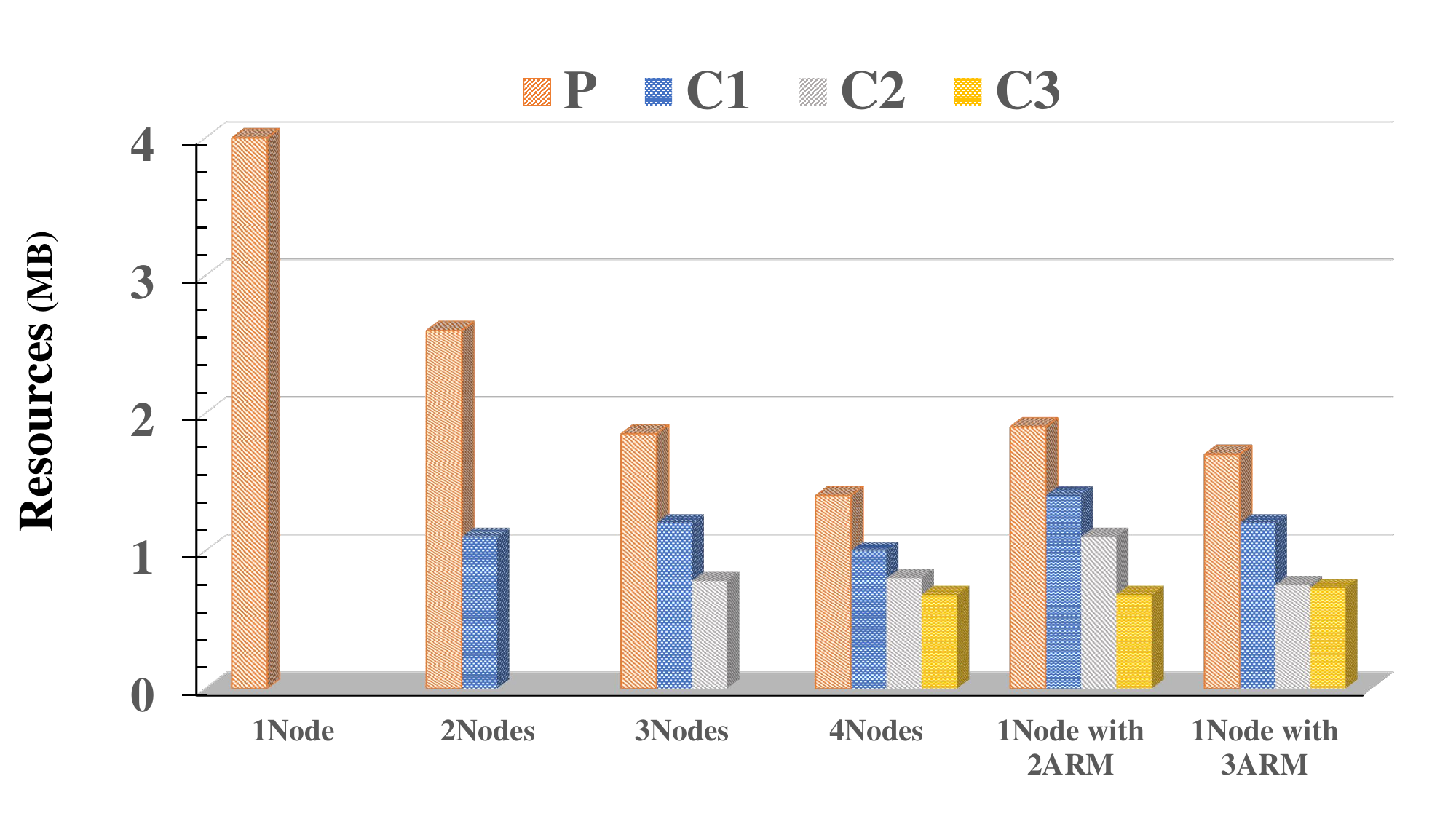}
\vspace{-2.5em}
\caption{Resource Consumption for Inference Over n Nodes} 
\label{fig:resources}
\vspace{-1.5em}
\end{figure}

\subsection{Impact of Proposed Technique on Resource Consumption}
\revision{The resource consumption of different worker nodes can be observed in Figure \ref{fig:resources}. 
In Figure \ref{fig:resources}, P represents the parent node, C1 represents the first child node, C2 represents the second child node and C3 represents the third child node. 
We observe the resource consumption for the following scenarios: a) the parent node has sufficient resources; b) the parent node does not have enough resources and outsources to multiple nodes.
The inference task takes 4 MB of data to complete. 
In the first case, the single parent node P can provide this data to complete the inference task. In other cases, the inference task is divided between multiple nodes (model parallelism), so the data required is also divided into multiple nodes. 
We can observe that the resources are not equally consumed in the parent and child nodes. This is because the parent node usually has additional steps to perform, such as the gradient collection from child nodes, and integrating them, thus, it consumes high resources. 
When compared to using a single node, in model parallelism, the required resources are provided from various nodes, and this helps to improve the processing speed.} 

\section{Conclusion}
\label{conclusion}

We proposed a resource- and workload-aware model parallelism-based malware detection technique which employs distributed training to enable better security for resource-constrained IoT devices. The metrics of distributed training on multiple nodes are analyzed and a speed-up of 9.8$ \times$ is observed compared to on-device training. From the results presented, it is also evident that the proposed technique produces state-of-the-art malware detection accuracy of 96.7\% among HPC-based detection techniques. 
Thus, the proposed technique is reliable and accurate for malware detection in IoT mid-scale devices.
\vspace{-1em}

\bibliographystyle{IEEEtran}
\bibliography{reference.bib}


\end{document}